\documentclass[preprint,12pt]{elsarticle}

\usepackage{graphicx}
\usepackage{amssymb}
\usepackage{lineno}

\journal{Fluid Phase Equilibria}

\begin{document}

\begin{frontmatter}


\title{Determination of pressure-viscosity relation of 2,2,4-trimethylhexane by all-atom molecular dynamics simulations}



\author[1]{Marco A. Galvani Cunha\corref{cor1}}
\author[1]{Mark O. Robbins}

\address[1]{Department of Physics \& Astronomy, Johns Hopkins University, Baltimore, MD, 21218, United States}

\cortext[cor1]{Corresponding author. E-mail: mgalvan1@jhu.edu}

\begin{abstract}
The Newtonian viscosity of 2,2,4-trimethylhexane at 293K is determined at pressures from 0.1MPa to 1000MPa.
Non-equilibrium molecular dynamics simulations are performed using AIREBO-M, an all-atom potential for hydrocarbons especially parameterized for high pressures.
The steady-state shear stress and viscosity are determined from simple shear simulations at rates between $10^7$ and $5\cdot 10^9\ \textrm{s}^{-1}$.
At low pressures, simulation rates are low enough to reach the Newtonian regime.
At high pressures, results are extrapolated to the Newtonian limit by fitting rate-dependent viscosities to Eyring theory.
The resulting pressure dependent viscosity is typical of small molecules and fits to a common model are discussed.
\end{abstract}


\end{frontmatter}


\section{Introduction}
\begin{figure}[bt]
\centering
\includegraphics[width=.6\textwidth,scale=1]{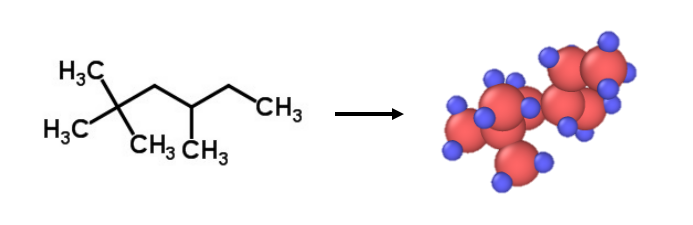}
\caption{2,2,4-trimethylhexane molecule and its representation in our all-atom simulations. Carbons are colored red and hydrogens are colored blue. The sphere sizes are not representative of the interaction potential.}
\label{molecule}
\end{figure}

The 10th Industrial Fluid Properties Simulation Challenge tasked simulators with predicting the pressure dependence of the Newtonian viscosity $\eta_N$ of a simple hydrocarbon liquid using any molecular modeling method. The hydrocarbon chosen was 2,2,4-trimethylhexane (Figure \ref{molecule}), and its viscosities were measured at pressures of 0.1, 25, 50, 100, 150, 250, 400, 500, 600, 700, 800, 900 and 1000 MPa. Here we present calculations of the viscosities at these pressures using all-atom nonequilibrium molecular dynamics simulations.
This paper was originally submitted as an entry to the challenge and won.
After the release of the results we added the benchmark data to our plot of the pressure-viscosity relation, Fig. \ref{pvisc}, and added a discussion of the
comparison between the results to the discussion. The rest of the paper describes the approach and remains unchanged unless noted.

This challenge was timely because the pressure dependence of fluid viscosities
plays a critical role in elastohydrodynamic lubrication (EHL)
\cite{dowson_elasto-hydrodynamic_1977,vergne_classical_2014,bair_classical_2016} and faster computers are
enabling calculation of fluid response at strain rates approaching experiments
for the first time \cite{dini2017,jadhao_probing_2017}.
In addition there has been an active debate about the connection between
molecular-scale processes and the rate and pressure dependence of viscosity
\cite{spikes.tbl.2014,bair.tbl.2015,spikes.tbl.2015}.
The challenge results will be helpful in determining which molecular interaction potentials and protocols are able to capture experimental behavior
\cite{jadhao_probing_2017,dini2016}.

Here we follow the recent study of squalane by Jadhao and Robbins \cite{jadhao_probing_2017}.
Nonequilibrium molecular dynamics simulations of simple shear are performed
as a function of strain rate $\dot{\gamma}$ at each pressure $p$.
At low pressures, simulation rates are low enough to reach the Newtonian regime,
and $\eta_N$ is evaluated directly from the calculated limiting viscosity.
As $p$ and $\eta_N$ rise, relaxation times become longer than our simulation
times, and we can not reach the Newtonian regime.
Values of $\eta_N$ for these high pressures are obtained by fitting the
rate-dependent shear stress to the Eyring model \cite{eyring_viscosity_1936}. The Eyring model assumes that flow occurs via thermally activated hops over an energy barrier that decreases linearly with the applied shear stress $\sigma$.
Combining the
probabilities of forward and backward hops leads to the Eyring equation relating strain rate and stress:
\begin{equation} \label{eq:eyring}
	\dot{\gamma} = \frac{\sigma_E}{\eta_N}\sinh{(\sigma/\sigma_E)} \ \ ,
\end{equation}
where the Eyring stress $\sigma_E$ is related to the sensitivity of the energy barrier to shear stress.
At large stresses $\sigma \sim \log{\dot{\gamma}}$ and
both parameters in the Eyring model, $\eta_N$ and $\sigma_E$,
can be obtained from a simple straight-line fit to a linear-log plot.

\section{Methods}

Non-equilibrium molecular dynamics simulations have long been used to determine thermal and transport properties of simple hydrocarbons \cite{allen_predicting_1997,liu_pressureviscosity_2015,allen_computer_2009}.
A key factor in the accuracy of the results is the choice of interaction
potential \cite{dini2016,allen_computer_2009}.
Our past work showed that united-atom and all-atom potentials gave similar results for the large molecule squalane \cite{jadhao_probing_2017} but an all-atom potential was required for the small molecule cyclohexane. We thus chose to use an all-atom potential for simulations of 2,2,4-trimethylhexane whose
geometry is shown in Figure \ref{molecule}. 

One common all-atom potential is the Adaptive Intermolecular Reactive Empirical Bond Order (AIREBO) potential \cite{stuart_reactive_2000}, which uses the reactive bond order (REBO) potential \cite{brenner_second-generation_2002} for intramolecular energies and a Lennard-Jones (LJ) potential for intermolecular interactions. It has been used successfully in many simulations of hydrocarbons at ambient pressure but over-predicts the stiffness of alkanes at high pressures due to the strong divergence in the LJ potential. O'Connor \textit{et al.} \cite{oconnor_airebo-m:_2015} developed a modified version of AIREBO called AIREBO-M that substitutes the LJ intermolecular interaction with a Morse potential. This softens the repulsive region, leading to better agreement with experimental data at high pressures. The potential was parameterized using x-ray data for the C-C separation in graphite and high-quality quantum chemistry calculations for C-H and H-H interactions. It was then validated against shock Hugoniot and crystal structure data for polyethylene at pressures up to 40GPa.

We determined the Newtonian viscosity of 2,2,4-trimethylhexane using 
non-equilibrium molecular dynamics (NEMD) simulations with AIREBO-M 
and the protocol used by Jadhao and Robbins to determine the viscosity of squalane
\cite{jadhao_probing_2017}.
All simulations were performed with the molecular dynamics package LAMMPS using standard protocols \cite{plimpton_fast_nodate}. The simulations were initialized by placing 1000 randomly oriented molecules of 2,2,4-trimethylhexane far apart from each other in a simulation box with periodic boundary conditions.
The box was then compressed slowly to reach the experimental density at room temperature and pressure and
this initial state was equilibrated at a temperature of 293K for 2.5ns using a Nos\'e-Hoover thermostat. The system was then brought to the desired pressure and equilibrated over 5ns using a Nos\'e-Hoover barostat. Different protocols for equilibration and changes in pressure and temperature gave equivalent results within statistical error bars. The timestep used for results below was 0.5fs, and the thermostat and barostat time constants were 0.5ps and 5.0ps, respectively. Both time constants are longer than the relevant correlation times in the system.

Simple shear simulations were done at constant density using standard NEMD methods. We imposed a shear velocity profile by deforming the periodic simulation cell at a constant strain rate and integrating the SLLOD equations of motion \cite{evans_nonlinear-response_1984}. We determined the shear stress by calculating the off-diagonal component of the stress tensor during the run, and monitored relevant state variables (temperature, pressure and shear stress) to determine when the system reached steady-state. Strains of order 10 were used to gather enough independent samples in steady-state for most pressures and strain rates. Only rates up to $5 \times 10^{9}\ \textrm{s}^{-1}$ were considered because heating and other nonequilibrium effects were
noticeable for squalane at $10^{10}\ \textrm{s}^{-1}$ \cite{jadhao_probing_2017} and the goal of the challenge was to determine the low rate viscous response.

During each simulation we acquired values for the components of the stress tensor averaged over successive time intervals of 5ps. After reaching steady-state, these data were averaged to obtain the mean shear stress and pressure. Since correlation times change with pressure and strain rate, care is needed in determining the statistical uncertainty in the average.  
For each run, a block time-averaging method is used \cite{frenkel_understanding_2002}. The variance of the distribution of block averages is calculated as a function of block size. Initially the variance rises with block size since sequential samples are correlated, but after the block size exceeds the correlation time the variance saturates to the true error on the mean. 
For the one run at the lowest rate, $10^7\ \textrm{s}^{-1}$, the simulation ran for a strain less than unity and may not have reached steady state. This introduces an additional systematic uncertainty that was estimated as about twice the uncertainty from block averages by looking at data from higher rates.

All runs are done at fixed density and the corresponding pressure may differ slightly from that set in the equilibration run, particularly at high pressures where the relaxation times are long. The pressure for the chosen density was determined by extrapolating $p(\dot{\gamma})$ to zero strain rate. In the Newtonian regime $p(\dot{\gamma})$ rises as $\dot{\gamma}^2$ and fits to this form are used to extrapolate to zero strain rate. For p = 800MPa and 900MPa we did not reach this regime before the deadline for the challenge, so our error bars were wide and were based on extrapolating the lowest strain rate data using the observed variation of pressure with rate for 1000MPa where lower rates were studied.
Adding the additional data shown in Fig. \ref{stressvisc} did not lead to significant changes in the extrapolated pressure, although the error bars decreased slightly. We have chosen not to change the table but the best fit values for 800 and 900 MPa would increase by about 5MPa and decrease by 10MPa, respectively.

\section{Results}

\begin{figure}[ht]
\centering
\includegraphics[width=.48\textwidth,scale=1]{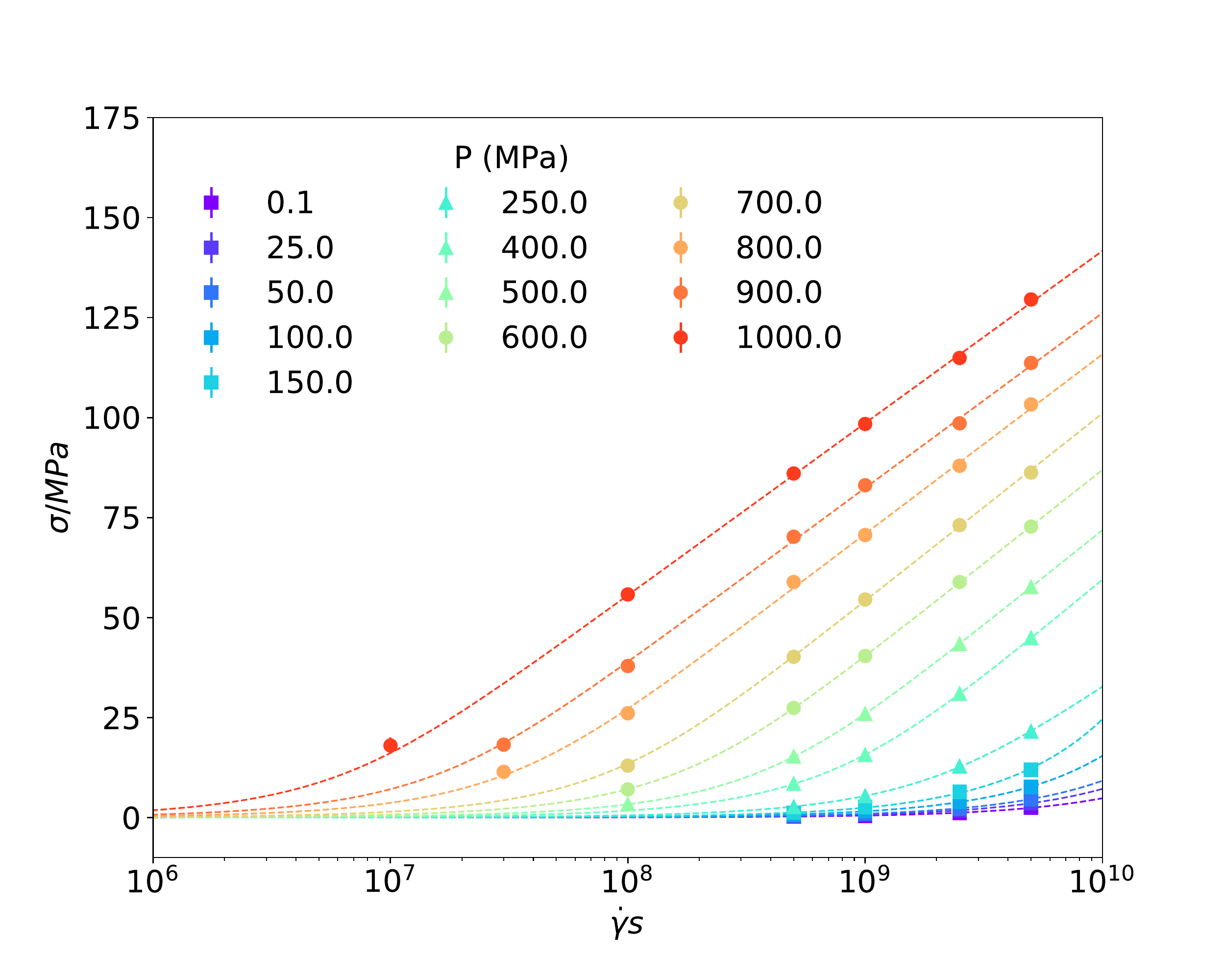}
\includegraphics[width=.48\textwidth,scale=1]{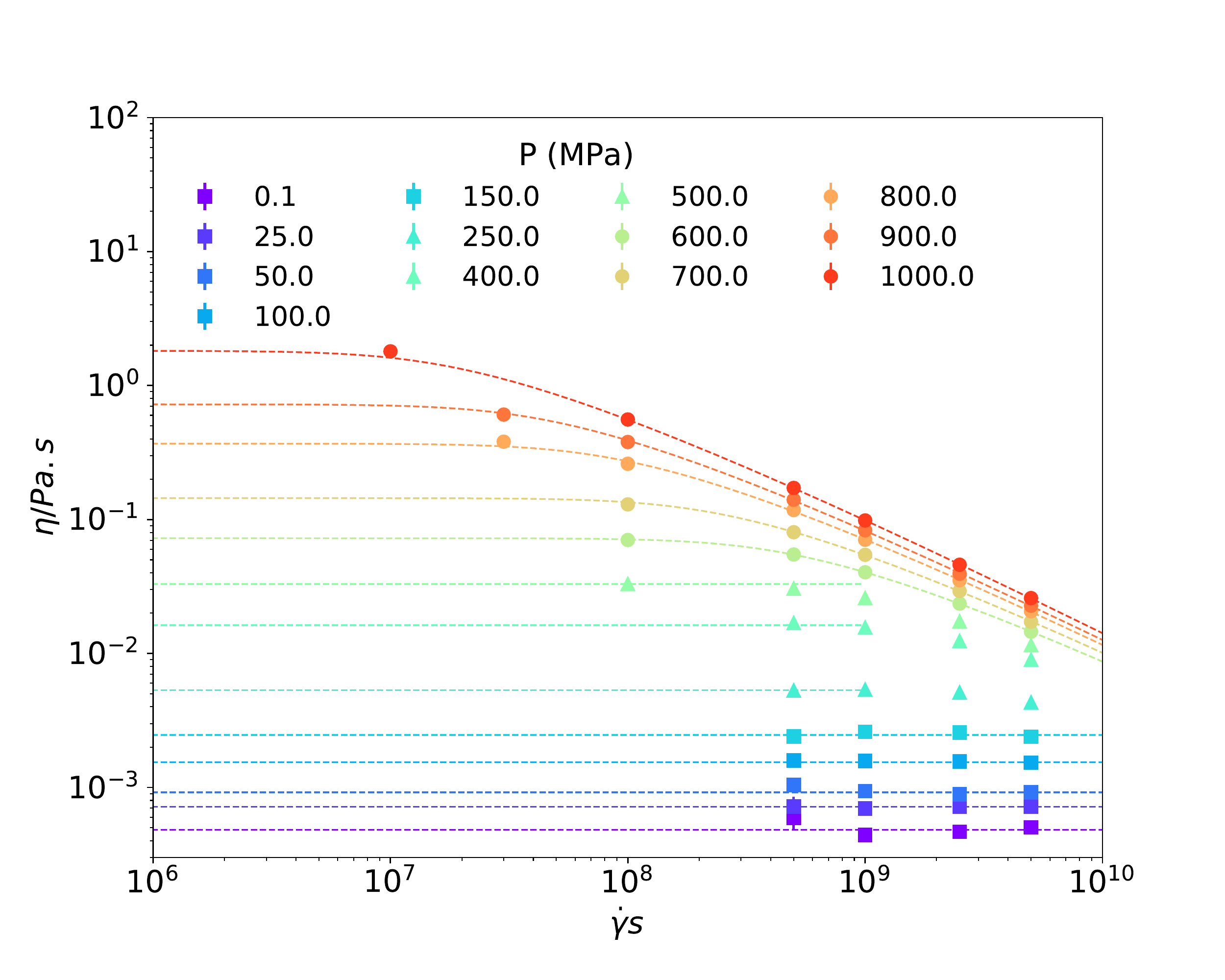}
\caption{(a) Stress and (b) viscosity as a function of strain rate at $T=293K$ and the indicated pressures.
Statistical error bars are shown when they are larger than the symbol size.
Results for each pressure are indicated by different colors and higher curves are for higher pressures in each panel.
Symbol types are used to indicate how $\eta_N$ was obtained.
Squares indicate that all data were in the Newtonian regime and the reported
$\eta_N$ is a weighted average over all data.
Triangles indicate pressures where the lowest two rates had reached the Newtonian limit and were averaged to determine $\eta_N$.
Horizontal dashed lines in panel (b) indicate the average viscosity obtained for these
pressures (squares and triangles).
Eyring fits were used to determine $\eta_N$ for the higher pressures indicated by circles.
These fits are shown for $p\geq 600$MPa in panel (b) and for all pressures in
panel (a), where they should only be considered guides to the eye for $p \leq 500$MPa.
Data at strain rates of $3 \cdot 10^7\ \textrm{s}^{-1}$ and $2.5 \cdot 10^9\ \textrm{s}^{-1}$ were obtained after the contest submission.
They were not used to refit the lines, because they are within statistical errors of the original fits.
}
\label{stressvisc}
\end{figure}

Figure \ref{stressvisc} shows results for the shear stress $\sigma$ and nonequilibrium viscosity $\eta(\dot{\gamma}) = \sigma/\dot{\gamma}$ as a function of shear rate for each pressure.
As noted in our original submission, further data was being taken, but not available at the time of the submission. The new points are at strain rates of $3 \cdot 10^7\ \textrm{s}^{-1}$ and $2.5 \cdot 10^9\ \textrm{s}^{-1}$ and lie close to the fit lines which were obtained without them.
As a result, refitting led to changes well within our
error bars and we report the original results for $\eta_N$ below.

At pressures less than 250MPa there is no statistically significant change in viscosity at the rates studied, indicating that relaxation times are less than 0.1ns.
As $p$ increases, shear thinning becomes more pronounced, and sets in at progressively lower rates.
For $p \leq 500$MPa the lowest simulation rates appear to reach a plateau corresponding to the Newtonian viscosity.
At higher pressure, $\eta$ increases significantly between the lowest two shear rates.
%Further data is being obtained at lower rates, but is not available at this submission deadline.

Values of $\eta_N$ were obtained from the above simulation data following the procedure of Ref. \cite{jadhao_probing_2017}.
For $p \leq 500$MPa the Newtonian viscosity in Table \ref{tab:sim}
was obtained from an average over points in the plateau weighted by the
statistical uncertainty obtained as described in the previous
section.
Squares in Fig. \ref{stressvisc} indicate pressures where all rates were used in the average.
Triangles indicate pressures where the lowest two rates were averaged.
Horizontal dashed lines in Fig. \ref{stressvisc}(b) show the Newtonian viscosity
obtained from these averages.

The dashed lines at higher pressures in Fig. \ref{stressvisc}(b) and at all pressures in Fig. \ref{stressvisc}(a) are fits to Eyring theory.
For $p \leq 500$MPa they should be considered as guides to the eye
because the range of shear thinning is not sufficient
to distinguish between different models of shear thinning and
earlier work has shown that Eyring theory only applies at high pressures \cite{jadhao_probing_2017}. 

Eyring theory provides a good fit to data for 2,2,4-trimethylhexane at $ p \geq 600$Mpa.
The slope in plots of $\sigma$ against the logarithm of strain rate
is proportional to the Eyring stress $\sigma_E$.
As for squalane\cite{jadhao_probing_2017}, $\sigma_E$ is nearly independent of pressure
at fixed temperature and the fits for $p \geq 600$MPa all give
$\sigma_E = 20 \pm 1$MPa.
In Eyring theory, $\sigma_E = k_B T/V^*$, where $V^*$ is called the activation volume and represents the sensitivity of the energy barrier to stress.
While $V^*$ need not correspond to any actual volume, fits frequently give values comparable to the molecular volume. 
The fits in Fig. \ref{stressvisc} correspond to $V^* \approx 0.2$nm$^{3}$ which is close to the molecular volume of $0.29$nm$^{3}$.

Eyring fits were used to determine the values of $\eta_N$ in Table \ref{tab:sim}
for $p \geq 600$MPa.
Because this represents an extrapolation, the error in $\eta_N$ is
more difficult to estimate.
The quoted errors are the larger of the uncertainty in $\eta$ at the
lowest $\dot{\gamma}$ and the calculated uncertainty in the fit.
The ratio of the extrapolated $\eta_N$ to the largest calculated value
was up to a factor of 1.7 for $p=800$ and $900$MPa.

\begin{figure}[t]
\centering
\includegraphics[width=.9\textwidth,scale=1]{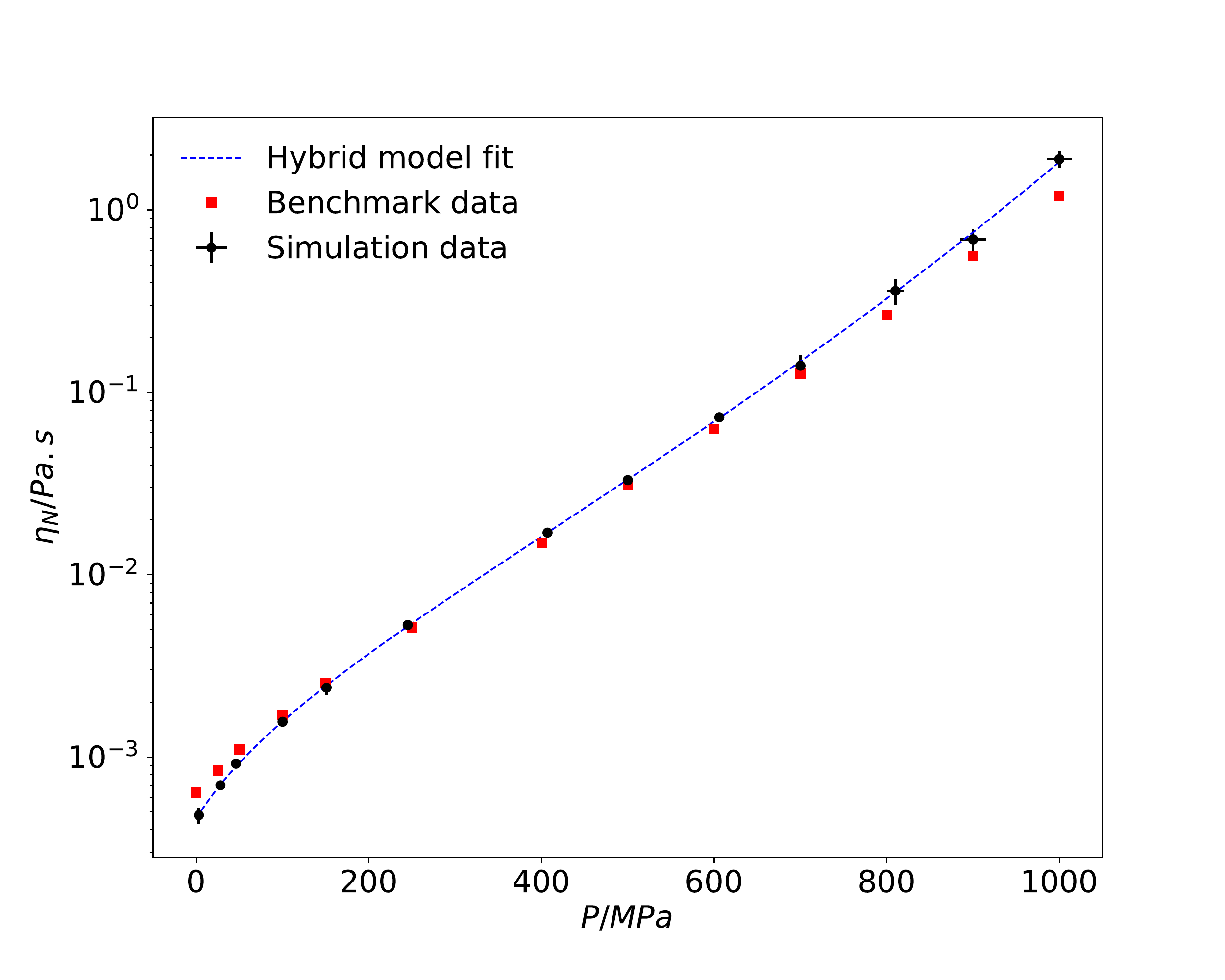}
\caption{Pressure-viscosity relation for 2,2,4-trimethylhexane. Simulation data is from Table \protect\ref{tab:sim}, and statistical error bars are shown when larger than the symbol size. The dashed line is a fit to Eq. \protect\ref{eq:hybrid} with coefficients $\eta_0 = 0.46 \pm 0.02$mPa-s, $C_F = 14 \pm 4$, $p_\infty = 3.5 \pm 0.6$GPa, $a_0 = 0.013 \pm 0.002$MPa$^{-1}$, $q = 0.8 \pm 0.2$.}
\label{pvisc}
\end{figure}

Figure \ref{pvisc} shows the Newtonian viscosity as a function of pressure from Table \ref{tab:sim}.
Note that there is an inflection point in the data at a pressure of around 500 to 600MPa.
This type of curve is commonly observed in experimental plots of $\log \eta_N$ against pressure and procedures have been developed for fitting it.

Paluch \cite{paluch_scaling_1999} proposed a pressure-viscosity relation that assumes the existence of a finite pressure $p_\infty$ at which relaxation times diverge at a given temperature. It is analogous to the temperature where $\eta_N$ diverges in a Vogel-Fulcher-Tammann (VFT) \cite{vogel,tammann_abhangigkeit_1926,fulcher_analysis_1925} theory for the glass transition. McEwen \cite{mcewen} proposed a relation that explained the slower than exponential rise with pressure seen at low $p$.
Bair \cite{bair_hybrid} combined the Paluch and McEwen equations into a single relation called the hybrid model:
\begin{equation}
	\eta = \eta_0\exp{\Bigg(\frac{C_Fp}{p_\infty-p}\Bigg)}(1+a_0 p/q)^q
    \label{eq:hybrid}
\end{equation}
The dotted line in Fig. \ref{pvisc} is a fit of Equation \ref{eq:hybrid} with values given in the caption.
As with many experiments, Equation \ref{eq:hybrid} provides a good fit to the data, but there were substantial uncertainties in the fit values given the large number of parameters.

The challenge asked for the viscosity at precise values of pressure and our simulations were at slightly different values.
Table \ref{tab:challenge} gives interpolated values of $\eta_N$ at the challenge pressures.
The first column was obtained from a linear interpolation between neighboring data points from our simulations.
The shifts are no larger than the errorbars in Table \ref{tab:sim}.
The second column was obtained from the fit to Equation \ref{eq:hybrid} shown in Figure \ref{pvisc}.
The two columns are consistent within the statistical errors of
Table \ref{tab:sim} and $\eta_N^{lin}$ represents our official entry.

\begin{table}[t]
\centering
\begin{tabular}[t]{c|c|c}
\hline
\hline
\boldmath$\rho (kg/m^3)$ & \boldmath$p(MPa)$ & \boldmath$\eta_N(mPa.s)$\\
\hline
710 & 3.0 $\pm$ 0.2 & 0.48 $\pm$ 0.05\\
742 & 28 $\pm$ 1 & 0.70 $\pm$ 0.03\\
758 & 46 $\pm$ 1 & 0.92 $\pm$ 0.05\\
791 & 100 $\pm$ 1 & 1.56 $\pm$ 0.04\\
814 & 151 $\pm$ 1 & 2.4 $\pm$ 0.2\\
846 & 245 $\pm$ 1 & 5.3 $\pm$ 0.3\\
885 & 407 $\pm$ 3 & 17 $\pm$ 1\\
903 & 500 $\pm$ 2 & 33 $\pm$ 2\\
921 & 606 $\pm$ 2 & 73 $\pm$ 3\\
935 & 700 $\pm$ 2 & 140 $\pm$ 20\\
950 & 810 $\pm$ 10 & 360 $\pm$ 60\\
960 & 900 $\pm$ 15 & 690 $\pm$ 100\\
974 & 1000 $\pm$ 15 & 1900 $\pm$ 200\\
\hline  % Please only put a hline at the end of the table
\end{tabular}
\caption{Nonequilibrium simulations were done
	as a function of rate at the indicated densities $\rho$.
Values for the corresponding pressure and viscosity in the Newtonian limit
were obtained as described in the text.
The errorbar for $P=1000MPa$ was originally submitted without the stated factor of two in the methods and is corrected here.
}
\label{tab:sim}

\end{table}
\begin{table}
\centering
\begin{tabular}[t]{c|c|c}
\hline
\hline
\boldmath$p(MPa)$ & \boldmath$\eta_N^{lin}(mPa.s)$ & \boldmath$\eta_N^{hyb}(mPa.s)$\\\hline
0.1 &  0.46 $\pm$ 0.05 &0.47\\
25 &  0.67 $\pm$ 0.03 & 0.69\\
50 &  0.96 $\pm$ 0.05 & 0.94\\
100 & 1.56 $\pm$ 0.04 & 1.6\\
150 &  2.4 $\pm$ 0.2 & 2.4\\
250 &  5.5 $\pm$ 0.3 & 5.4\\
400 & 16 $\pm$ 1 & 16\\
500 & 33 $\pm$ 2 & 33\\
600 & 70 $\pm$ 3 & 69\\
700 & 140 $\pm$ 20 & 148\\
800 & 330 $\pm$ 60 & 328\\
900 &  690 $\pm$ 100 & 755 \\
1000 & 1900 $\pm$ 200 & 1827\\
\hline  % Please only put a hline at the end of the table
\end{tabular}
\caption{Pressures specified in the challenge and corresponding Newtonian viscosities obtained by linear interpolation, $\eta_N^{lin}$, and fits to Eq. 2, $\eta_N^{hyb}$.
}
\label{tab:challenge}

\end{table}

\section{Discussion}

The results presented above show that all-atom NEMD simulations can readily reach low enough rates to be in the Newtonian limit for $\eta_N$ as large as 0.1Pa.s.
With sufficient time, rates of $10^7\ \textrm{s}^{-1}$ and viscosities of 1Pa.s can
be reached.
The computational effort grows linearly with viscosity for both nonequilibrium
simulations and equilibrium methods based on the fluctuation-dissipation
theorem \cite{allen_computer_2009}.
Thus calculations of viscosities much larger than 1Pa.s require a different approach.
In this paper and Ref. \cite{jadhao_probing_2017}, Eyring theory was used to extrapolate high-rate data to the Newtonian limit.
For squalane the results were consistent with trends in experimental data over a wide range of pressure and temperature.

For 2,2,4-trimethylhexane the combined approach of direct evaulation and Eyring extrapolation gave values of $\eta_N$ with numerical uncertainties of about 15\% or less.
The uncertainties associated with the choice of atomic force field are likely to be much larger.
Ewen et al. \cite{dini2016} calculated $\eta_N$ for n-hexadecane with
all-atom and united-atom potentials and found
values a factor of 2 lower or higher than experiment.
O'Connor \cite{oconnorthesis} found a variation by $\pm 50$\% in the friction between polyethylene chains calculated with AIREBO-M and other all-atom potentials.
Based on these past studies, we expect the systematic error associated
with the choice of potential to be as large as a factor of 2.

The goal of the challenge is to determine how well viscosities can be predicted,
so we made no attempt to modify the AIREBO-M force field.
It was parameterized based on structural data at high pressures and is expected
to be least accurate at ambient pressure and high temperature \cite{oconnor_airebo-m:_2015}.
More accurate results might be obtained by refitting AIREBO-M to match
the known viscosities of molecules similar to the target molecule.

To assess the accuracy of AIREBO-M before the challenge results were announced we compared the results in Table \ref{tab:challenge} to published experiments.
The calculated viscosities have roughly the same magnitude as similar hydrocarbons at lower pressures \cite{dymond_transport_nodate-1,dymond_transport_nodate-2,oliveira_viscosity_1992}.
Johnson and Fawcett \cite{johnson_inter-polymerization_1946} measured the viscosity of 2,2,4-trimethylhexane at room temperature and pressure and found a viscosity of $0.648$mPa.s.
This is higher than our ambient pressure value by about 40\% and at a
slightly higher temperature,
suggesting that our values might be systematically low by up to 50\%.
As noted, AIREBO-M was optimized for high pressure.
As a check of high pressure accuracy, we used the same model
to simulate simple shear of 2,2,4-trimethylpentane, also known as isooctane, at $T = 298K, p = 500MPa$.
The simulations gave a viscosity of $15.3\pm 0.5$mPa.s, while
Dymond \cite{dymond_transport_nodate} reports a measured value that is about
10\% smaller, $13.6$mPa.s with an uncertainty of $2\%$.
This is consistent with an increase in the accuracy of AIREBO-M at
high pressure.

%\textcolor{red}{We look forward to finding out the actual success of the method when the results of the challenge are announced}.
After the challenge data were announced, we added them to Fig. \ref{pvisc} to facilitate comparison.
%k\textcolor{blue}{The same trends can also be seen when comparing the simulation values with the experimental benchmark in \ref{pvisc}.
For all the pressures simulated, the predicted viscosities differ from experiment by less than a factor of 2.
This is consistent with our estimate above of the systematic errors associated with the choice of the potential.
Calculated viscosities are in excellent agreement with experiment at intermediate pressures, with the largest deviations at low and high pressures that we now discuss.

The difference in $\eta_N$ at ambient conditions is slightly better than expected because the challenge data are lower than previously published values \cite{johnson_inter-polymerization_1946}.
Some challenge entries were more accurate in this limit because they fit to it or
used potentials developed for ambient pressure.
Our underestimate of $\eta_N$ in ambient conditions is consistent with known behavior of
AIREBO \cite{Stuart}, which is nearly the same as AIREBO-M at ambient pressure.
AIREBO gives densities that are too large because it underestimates the attractive cohesive potential. An improved version \cite{Stuart} that uses chemistry dependent van der Waals interactions and includes long-range tail corrections to the pressure \cite{frenkel_understanding_2002} provides better densities at ambient pressure but the potential is more complicated and not available through LAMMPS. These van der Waals terms become increasingly irrelevant as pressure increases.

Our predictions capture the viscosity-pressure coefficient and the location of the inflection point in $\eta_N$ quite accurately but rise slightly too rapidly at the highest pressures.
Based on other entries to the challenge, this rise seems to be very sensitive to the potential.
Some entries saw no inflection and others found much steeper rises than in our work.
For our data the largest deviation ($\sim~50$\%) is at 1000MPa.
Note that a single simulation at the lowest rate was most critical in determining this value and, as noted, this run may not have gone to strains large enough to reach steady state. This typically led to an overestimate of viscosity for squalane \cite{jadhao_probing_2017} and it is possible that additional low rate data would produce values closer to experiment.
It would be very interesting to explore still higher pressures with both experiment and simulation to test the hybrid model and predicted divergence of $\eta$ and to see what is needed to improve interaction potentials.

\section{Acknowledgements}
We thank Vikram Jadhao for useful discussions.
Funding: This material is based upon work supported by the National Science Foundation under Grant No. DMR-1411144; and the Army Research Laboratory under the MEDE Collaborative Research Alliance, through Grant W911NF-12-2-0022.

%% References


%% \bibliographystyle{model1-num-names}
%% \bibliography{tmh.bib}

\begin{thebibliography}{10}

\bibitem{dowson_elasto-hydrodynamic_1977}
D.~Dowson and G.~R. Higginson, {\em Elasto-hydrodynamic lubrication}.
\newblock International series on materials science and technology ; v. 23,
  Oxford [Eng.] ; New York: Pergamon Press, si ed~ed., 1977.

\bibitem{vergne_classical_2014}
P.~Vergne and S.~Bair, ``Classical {EHL} {Versus} {Quantitative} {EHL}: {A}
  {Perspective} {Part} {I}—{Real} {Viscosity}-{Pressure} {Dependence} and the
  {Viscosity}-{Pressure} {Coefficient} for {Predicting} {Film} {Thickness},''
  {\em Tribology Letters}, vol.~54, pp.~1--12, Apr. 2014.

\bibitem{bair_classical_2016}
S.~Bair, L.~Martinie, and P.~Vergne, ``Classical {EHL} {Versus} {Quantitative}
  {EHL}: {A} {Perspective} {Part} {II}—{Super}-{Arrhenius} {Piezoviscosity},
  an {Essential} {Component} of {Elastohydrodynamic} {Friction} {Missing} from
  {Classical} {EHL},'' {\em Tribology Letters}, vol.~63, Sept. 2016.

\bibitem{dini2017}
J.~P. Ewen, C.~Gattinoni, J.~Zhang, D.~M. Heyes, H.~A. Spikes, and D.~Dini,
  ``On the effect of confined fluid molecular structure on nonequilibrium phase
  behaviour and friction,'' {\em Phys. Chem. Chem. Phys.}, vol.~19,
  pp.~17883--17894, 2017.

\bibitem{jadhao_probing_2017}
V.~Jadhao and M.~O. Robbins, ``Probing large viscosities in glass-formers with
  nonequilibrium simulations,'' {\em Proceedings of the National Academy of
  Sciences}, vol.~114, pp.~7952--7957, July 2017.

\bibitem{spikes.tbl.2014}
H.~A. Spikes and Z.~Jie, ``History, origins and prediction of
  elastohydrodynamic friction,'' {\em Tribology Letters}, vol.~56, no.~1,
  pp.~1--25, 2014.

\bibitem{bair.tbl.2015}
S.~Bair, P.~Vergne, P.~Kumar, G.~Poll, I.~Krupka, M.~Hartl, W.~Habchi, and
  R.~Larsson, ``Comment on “history, origins and prediction of
  elastohydrodynamic friction” by spikes and jie,'' {\em Tribology Letters},
  vol.~58, no.~1, 2015.

\bibitem{spikes.tbl.2015}
H.~A. Spikes and Z.~Jie, ``Reply to the comment by scott bair, philippe vergne,
  punit kumar, gerhard poll, ivan krupka, martin hartl, wassim habchi, roland
  larson on “history, origins and prediction of elastohydrodynamic
  friction” by spikes and jie,'' {\em Tribology Letters}, vol.~58, no.~1,
  2015.

\bibitem{dini2016}
J.~P. Ewen, C.~Gattinoni, F.~M. Thakkar, N.~Morgan, H.~A. Spikes, and D.~Dini,
  ``A comparison of classical force-fields for molecular dynamics simulations
  of lubricants,'' {\em Materials}, vol.~9, no.~8, p.~651, 2016.

\bibitem{eyring_viscosity_1936}
H.~Eyring, ``Viscosity, {Plasticity}, and {Diffusion} as {Examples} of
  {Absolute} {Reaction} {Rates},'' {\em The Journal of Chemical Physics},
  vol.~4, pp.~283--291, Apr. 1936.

\bibitem{allen_predicting_1997}
W.~Allen and R.~L. Rowley, ``Predicting the viscosity of alkanes using
  nonequilibrium molecular dynamics: {Evaluation} of intermolecular potential
  models,'' {\em The Journal of Chemical Physics}, vol.~106, pp.~10273--10281,
  June 1997.

\bibitem{liu_pressureviscosity_2015}
P.~Liu, H.~Yu, N.~Ren, F.~E. Lockwood, and Q.~J. Wang, ``Pressure–{Viscosity}
  {Coefficient} of {Hydrocarbon} {Base} {Oil} through {Molecular} {Dynamics}
  {Simulations},'' {\em Tribology Letters}, vol.~60, Dec. 2015.

\bibitem{allen_computer_2009}
M.~P. Allen and D.~J. Tildesley, {\em Computer simulation of liquids}.
\newblock Oxford science publications, Oxford: Clarendon Press, 2009.

\bibitem{stuart_reactive_2000}
S.~J. Stuart, A.~B. Tutein, and J.~A. Harrison, ``A reactive potential for
  hydrocarbons with intermolecular interactions,'' {\em The Journal of Chemical
  Physics}, vol.~112, pp.~6472--6486, Apr. 2000.

\bibitem{brenner_second-generation_2002}
D.~W. Brenner, O.~A. Shenderova, J.~A. Harrison, S.~J. Stuart, B.~Ni, and S.~B.
  Sinnott, ``A second-generation reactive empirical bond order ({REBO})
  potential energy expression for hydrocarbons,'' {\em Journal of Physics:
  Condensed Matter}, vol.~14, pp.~783--802, Feb. 2002.

\bibitem{oconnor_airebo-m:_2015}
T.~C. O’Connor, J.~Andzelm, and M.~O. Robbins, ``{AIREBO}-{M}: {A} reactive
  model for hydrocarbons at extreme pressures,'' {\em The Journal of Chemical
  Physics}, vol.~142, p.~024903, Jan. 2015.

\bibitem{plimpton_fast_nodate}
S.~Plimpton, ``Fast {Parallel} {Algorithms} for {Short}–{Range} {Molecular}
  {Dynamics},'' {\em Journal of Computational Physics}, vol.~117, pp.~1--19,
  1995. http://lammps.sandia.gov.

\bibitem{evans_nonlinear-response_1984}
D.~J. Evans and G.~P. Morriss, ``Nonlinear-response theory for steady planar
  {Couette} flow,'' {\em Physical Review A}, vol.~30, pp.~1528--1530, Sept.
  1984.

\bibitem{frenkel_understanding_2002}
D.~Frenkel and B.~Smit, {\em Understanding molecular simulation from algorithms
  to applications}.
\newblock San Diego, CA: Academic Press, 2002.

\bibitem{paluch_scaling_1999}
M.~Paluch, Z.~Dendzik, and S.~J. Rzoska, ``Scaling of high-pressure viscosity
  data in low-molecular-weight glass-forming liquids,'' {\em Physical Review
  B}, vol.~60, pp.~2979--2982, Aug. 1999.

\bibitem{vogel}
H.~Vogel {\em Z. Phys.}, vol.~22, p.~645, 1921.

\bibitem{tammann_abhangigkeit_1926}
G.~Tammann and W.~Hesse, ``Die {Abhängigkeit} der {Viscosität} von der
  {Temperatur} bie unterkühlten {Flüssigkeiten},'' {\em Zeitschrift für
  anorganische und allgemeine Chemie}, vol.~156, pp.~245--257, Oct. 1926.

\bibitem{fulcher_analysis_1925}
G.~S. Fulcher, ``{ANALYSIS} {OF} {RECENT} {MEASUREMENTS} {OF} {THE} {VISCOSITY}
  {OF} {GLASSES},'' {\em Journal of the American Ceramic Society}, vol.~8,
  pp.~339--355, June 1925.

\bibitem{mcewen}
E.~McEwen, ``The effect of variation of viscosity with pressure on the
  load-carrying capacity of the oil film between gear teeth,'' {\em J. Inst.
  Petroleum}, vol.~38, no.~344-345, pp.~646--672, 1952.

\bibitem{bair_hybrid}
S.~Bair, ``Choosing pressure-viscosity relations.,'' {\em High Temp.-High
  Press.}, vol.~44, no.~6, pp.~415--428, 2015.

\bibitem{oconnorthesis}
T.~C.~O'Connor, ``The Nonlinear Mechanics and Rheology of Oriented Polymers,'' Ph. D. Thesis, Johns Hopkins University, 2018.

\bibitem{dymond_transport_nodate-1}
J.~H. Dymond, ``Transport {Properties} of {Nonelectrolyte} {Liquid} {Mixtures}
  {II}. {Viscosity} {Coefficients} for the n-{Hexane} + n-{Hexadecane} {System}
  at {Temperatures} from 25 to 100$^o$C at {Pressures} {Up} to the
  {Freezing} {Pressure} or 500 {MPa},'' {\em International Journal of
  Thermophysics}, vol.~1, no.~4, pp.~345--373, 1980.

\bibitem{dymond_transport_nodate-2}
J.~H. Dymond, J.~Robertson, and J.~D. Isdale, ``Transport {Properties} of
  {Nonelectrolyte} {Liquid} {Mixtures} {III}. {Viscosity} {Coefficients} for
  n-{Octane}, n-{Dodecane}, and {Equimolar} {Mixtures} of n-{Octane} +
  n-{Dodecane} and n-{Hexane} + n-{Dodecane} from 25 to 100$^o$C at
  {Pressures} {Up} to the {Freezing} {Pressure} or 500 {MPa},'' {\em
  International Journal of Thermophysics}, vol.~2, no.~2, pp.~133--154, 1981.

\bibitem{oliveira_viscosity_1992}
C.~M. B.~P. Oliveira and W.~A. Wakeham, ``The viscosity of five liquid
  hydrocarbons at pressures up to 250 {MPa},'' {\em International Journal of
  Thermophysics}, vol.~13, pp.~773--790, Sept. 1992.

\bibitem{johnson_inter-polymerization_1946}
G.~C. Johnson and F.~S. Fawcett, ``Inter-polymerization of {Isobutene} and
  2-{Methyl}-2-butene {Using} an {Alumina}-{Silica} {Catalyst}. {Composition}
  of the {Hydrogenated} {Nonenes},'' {\em Journal of the American Chemical
  Society}, vol.~68, pp.~1416--1419, Aug. 1946.

\bibitem{dymond_transport_nodate}
J.~H. Dymond, ``Transport {Properties} of {Nonelectrolyte} {Liquid} {Mixtures}
  {VII}. {Viscosity} {Coefficients} for {Isooctane} and for {Equimolar}
  {Mixtures} of {Isooctane} + n-{Octane} and {Isooctane} + n-{Dodecane} from 25
  to 100$^o$C at {Pressures} up to 500 {MPa} or to the {Freezing}
  {Pressure},'' {\em International Journal of Thermophysics}, vol.~6, no.~3,
  pp.~233--250, 1985.
  
\bibitem{Stuart}
A.~Liu and S.~J.~Stuart, ``Empirical bond-order potential for hydrocarbons:
Adaptive treatment of van der Waals interactions,'' {\em Journal of Computational Chemistry}, vol.~29, no.~4, pp.~601-611, 2008.

\end{thebibliography}






% \end{thebibliography}

% \section*{References}

\end{document}